# Optically Switchable Fluorescence Enhancement at Critical Interparticle Distances

Arda Gulucu[1] and Emre Ozan Polat[1,2,*]

[1]*UNAM – National Nanotechnology Research Center and Institute of Materials Science and Nanotechnology, Bilkent University, 06800, Ankara, Turkey*

[2]*Department of Physics, Bilkent University, 06800, Ankara, Turkey*

**Corresponding author: emre.polat@bilkent.edu.tr*

**Abstract**

Plasmonic nanostructures provide electric field localization to be used as a fluorescence enhancement tool for the closely located fluorophores. However, metallic structures exhibit non-radiative energy transfer at close proximity, which suppresses the boost in the photoluminescence spectrum due to inhomogeneous medium. Compensation to non-radiative losses is fundamentally restricted, therefore defining the critical interparticle distances, where the fluorescence enhancement is detectable hold utmost importance for device applications. In this work, we numerically identified the critical interparticle distances of a metal nanoparticle (MNP) and quantum emitters (QEs) with angstrom resolution by analyzing the interplay between quantum yield and non-radiative decay. By engaging a collimated light application on silver nanoparticle (AgNP) placed at a critical distance, we simulated an active fluorescence enhancement switch yielding observable 7-fold increase in fluorescence intensity. The provided free space simulation includes the complete response of AgNP with retardation and higher order multi-polar effects for which the previous analytical works fall short. While the model bridges the absorption and emission spectra via corresponding Stokes shift values and presents a general approach for the interaction of QEs and MNPs in Rayleigh regime, it can be extended to Mie regime for larger QEs and can be modified for dielectric device environment.

## Introduction

Photoluminescence (PL) enhancement arises from the interaction between plasmonic nanostructures[1] that support localized surface plasmon polaritons (LSPPs)[2] and QEs such as semiconductor quantum dots (QDs)[3], molecular systems[4], defects[5] or vacancies[6]. LSPPs confine and amplify electromagnetic fields at the nanoscale, surpassing the diffraction limit and thereby markedly increasing the excitation efficiency of nearby QEs. In addition, plasmonic modes accelerate the spontaneous emission of QEs by providing additional radiative and non-radiative decay pathways, increasing the local photonic density of states (LDOS)[7].

As a QE, semiconducting QDs, particularly CdSe/ZnS core-shell QDs (CSQDs) have gathered immense attention due to their size tunable optical properties[8], high quantum yields and sustainable photostability[9]. Their size-dependent emission, a direct result of quantum confinement, allows for precise control over the emitted wavelength, making them invaluable in display technologies[10]. CdSe/ZnS CSQDs exhibit a broad absorption spectrum that extends from the UV to the visible region[11]. The absorption onset is size-dependent, with smaller QDs absorbing at shorter wavelengths (higher energy) and larger QDs absorbing at longer wavelengths (lower energy)[12]. CdSe is a semiconductor characterized by a relatively small direct band gap (approximately 1.7–1.8 eV in its bulk form), which increases substantially in quantum-confined nanocrystals, enabling absorption in the visible spectral range[13]. On the other hand, ZnS is a wide band gap semiconductor, with a bulk band gap of approximately 3.6–3.9 eV[14]. When CdSe and ZnS are combined in a core–shell configuration, they generally form a Type-I heterostructure. In this arrangement, both the conduction band minimum (CBM) and the valence band maximum (VBM) of the narrower band gap material (CdSe) lie energetically within the band gap of the wider band gap material (ZnS). When a photon is absorbed by the CdSe/ZnS QD, the electron-hole pair (exciton) is primarily confined within the CdSe core. The ZnS shell acts as a barrier, effectively trapping the exciton in the core. Superimposed on CdSe core, the ZnS shell plays a crucial role in passivating the CdSe core, significantly enhancing the quantum yield and photostability of these nanocrystals. ZnS capping results in high quantum yields (QYs) of up to 50–80% which is essential for long-duration applications like biological imaging[15] and QD displays[16]. Therefore, the CdSe core serves as the active light-emitting center in the core–shell quantum dot, with its band gap determining the main absorption and emission characteristics while the ZnS shell acts as a passive protective layer, improving the luminescence efficiency and stability of the core by passivating surface defects and reducing nonradiative recombination pathways.

The PL emission of CdSe/ZnS QDs is narrow and symmetric, with a full width at half maximum (FWHM) typically around 20 – 40 nm[17]. The emission wavelength is tunable by varying the core size, ranging from ~500 nm (green) to ~ 650 nm (red) for CdSe cores[18], making them excellent candidates for a wide spectrum of optoelectronic and biomedical applications.

The Stokes shift present in the emission spectrum of CdSe/ZnS CSQDs is characteristic feature—especially in the case of fluorescence—in which the emitted photons exhibit a longer wavelength (corresponding to lower energy) compared to the absorbed photons. At the single particle level, this energy difference between the maxima of the absorption and emission spectra is mainly sourced from vibrational relaxation and changes in molecular geometry such as equilibrium bond lengths and angles based on Frank-Condon principle[19,20]. In the ensemble measurements, however, contribution of solvent reorganization to overall Stokes shift become important[21].

Nanophotonics research has seen a tremendous acceleration during the last two decades by the implementations of MNPs and metamaterials that are able to confine EM field in subwavelength scales[22]. Originally described by Purcell[23], modification of spontaneous emission from a QE can be implemented in device applications by using plasmonic nanoantenna structures[24,25]. However, the plasmonic losses in metals due to non-radiative decay of plasmons strictly limits the application promises of plasmonic structures[26] where fluorescence enhancement is important[27,28]. Although the compensation mechanisms have been demonstrated by using low loss/high gain materials[29] or geometries[30], inherent losses associated to the absorption in metals fundamentally limit the ultimate performance of closely located fluorophores by non-radiative energy transfer[31]. To that end, understanding the spatial interplay between losses and quantum yield holds utmost importance to be able to design and implement photonic devices benefiting from the detectable fluorescence enhancement. Working in the single particle level may reveal this key enabling property, however, experimentally probing the EM interactions on a single particle level without a pre-developed enhancement strategy could be restricted due to fluorescence blinking that majorly hinders the real-time detection[32]. On the other hand, ensemble QD measurements include the interference of parameters such as size distribution[33], solvent interaction[34] and clustering effects[35] prohibiting the analyze of the essential nature of the coupling. Therefore, performing numerical models that are capable of of revealing the nature of spatial and temporal light matter interaction for anisotropic materials in nonhomogeneous mediums is pivotal for our understanding and the design of the future experiments.

Distance dependence for spontaneous decay rate of a single dipole emitter close to a metallic nanoparticle first reported by Carminati et al.[36] which had not been explicitly put forward in the previous theoretical fluorescent enhancement studies that consider Purcell effect[37–40]. The study reported that at short interparticle distances, the non-radiative decay rate follows an $R^{-6}$ dependence, akin to Förster resonance energy transfer. In contrast, the radiative decay rate displayed a more complex behavior, incorporating both $R^{-3}$ and $R^{-6}$ distance dependencies with a transition regime scaled with $R^{-4}$. However, the study only covers the simplest case as an analytical solution of the nanoparticle and emitter interacting through dipole-dipole coupling. Here in this study, we provide a comprehensive Finite Difference Time Domain (FDTD) model including MNP's complete electromagnetic (EM) response with retardation and higher-order multipolar effects and we report critical distance values for three different CdSe/ZnS CSQD sizes with angstrom resolution. To show the promises of critical positioning, we simulated an active optical control of fluorescence enhancement by the light excitation in a silver nanoparticle (AgNP). Our simulation treats the QEs as physical particles (CdSe/ZnS CSQDs) that absorb and scatter the incoming light and bridges their absorption spectra to their PL dipole emission by imposing corresponding Stokes shift values. While the approach presents a general free space model for QEs and MNPs in Rayleigh regime— where the particle sizes are much smaller than the incoming wavelength ($x \ll \frac{2\pi r}{\lambda}$)— the model can be extended to Mie regime for particles with larger sizes and further modified to be used in device structures implementing the corresponding environmental dielectric constants.

**Results**

To provide a complete light-matter interaction model of QEs, we obtained the absorption and scattering cross section spectra of CdSe/ZnS CSQDs under the illumination of an incoming light pulse. We employed three main CdSe core sizes with a constant ZnS shell size to identify and demonstrate the corresponding spectral responses. **Figure 1a-c** show the schematic illustration of the employed CSQDs as small dots with radius r = 2 nm (1 nm CdSe core + 1 nm ZnS shell), medium dots with radius r = 3 nm (2 nm core CdSe core + 1 nm ZnS shell) and large dots with radius r = 4nm (3 nm CdSe core + 1 nm ZnS shell). We irradiate the CSQDs by a broadband EM wave of 300 – 800 nm plane wave pulse with a 436nm peak position and 2 fs pulse length that is injected using total-field/scattered-field (TF/SF) method through Huygen's box source[41,42]. **Figure 1d** shows the obtained absorption cross section (ACS) spectrum. The absorption peaks shifts to the lower wavelengths (higher energies) with decreasing CSQD size. The left-hand side dominating peak (**Figure 1d**) in the ultraviolet region demonstrates an enhanced absorption at higher energies (shorter wavelengths) due to the increasing surface coverage of ZnS shell. While the constant thickness of ZnS shell (1 nm) does not exhibit a meaningful wavelength (energy) shift in UV absorption peak, increasing core CdSe results in a stark red shift in visible absorption peak. The resulting spectral parameters such as peak points, separation bandwidth and full width at half maximum (FWHM) values are given in the **Table 1**. Due to the tunable peak separation bandwidth, we emphasize that using the constant-shell-varying-core approach allows controllable band selection (rejection) for the photodiodes using CSQDs. To further investigate the EM interaction between the incoming irradiation and varying size CdSe/ZnS CSQDs, we obtained scattering cross section (SCS) spectrum.

**Figure 1e** shows SCS spectrum under the same broadband irradiation used for identifying the ACS spectrum. The profile of the SCS spectrum plays an important role in identifying the nature and efficiency of different scattering types such as Rayleigh, Mie and Raman scattering. For the employed CSQDs (with radius smaller than 10nm) the dominant scattering type is Rayleigh scattering since the particle size is much smaller than the incident wavelength. In Rayleigh scattering, the intensity is proportional to the sixth power of the particle radius while it is inversely proportional to the fourth power of the incident wavelength ($\sigma_{sca} \sim \frac{R^6}{\lambda^4}$). Therefore, the SCS spectrum in **Figure 1e** shows a dominating UV and blue light scatter compared to rest of the visible spectrum. The particle size increases the scattering intensity abruptly due to the sixth power dependence. Superposition of ACS (**Figure 1d)**, the SCS (**Figure 1e**) provides the profile of the extinction cross section that quantifies the resulting energy loss of the incident EM wave. The ACS intensities were found to be three orders of magnitude higher than the SCS intensities, therefore, superposition of the two spectrum will yield an extinction spectrum that is dominated by the ACS ($\sigma_{ext} =$ $\sigma_{sca} + \sigma_{abs} \approx \sigma_{abs}$) which is used to find the emission spectra of the CSQDs by adding the experimental Stokes shift values.

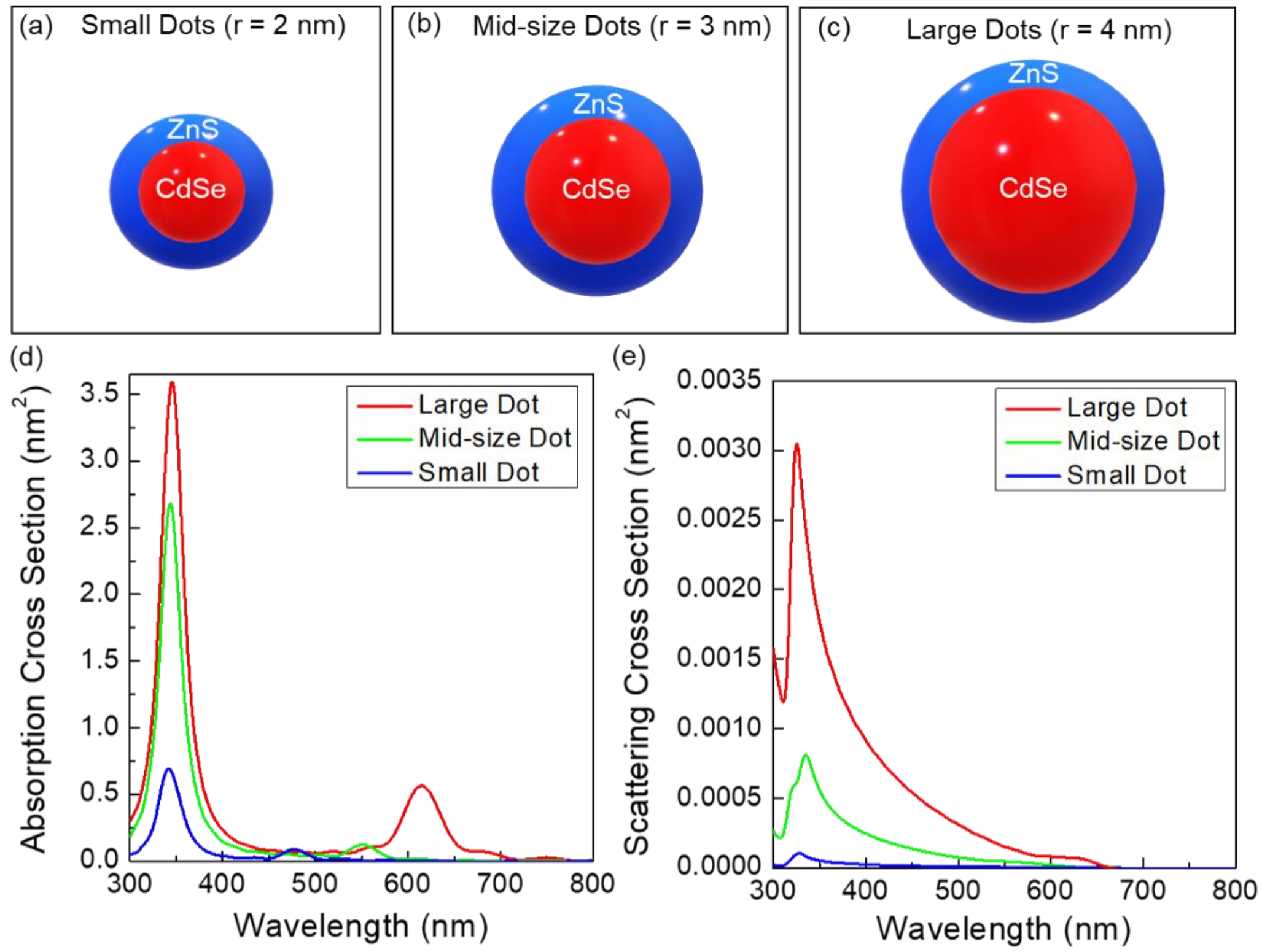

**Figure 1. (a)** Schematic illustration of CdSe/ZnS CSQDs employed in the FDTD simulations as r = 2 nm for the small dots, **(b)** r = 3 nm for mid-size dots and **(c)** r = 4 nm for the large dots. The core CdSe sizes are altered as $r_{core}$ = 1 nm, 2 nm and 3 nm while keeping the ZnS shell thickness constant as 1 nm. **(d)** Absorption cross section (ACS) spectra of large, mid-size and small dots under the irradiation of broadband EM wave pulse (300 – 800 nm) with a peak position of 436 nm. **(e)** Scattering cross section (SCS) spectra of large mid-size and small dots under the same illumination conditions that the ACS is obtained.

| ***CSQD Radius*** (nm) | ***UV Peak Point,*** $\lambda_{UV}$ (ZnS-Shell: 1 nm) | ***Visible Peak Point,*** $\lambda_{VIS}$ (CdSe-Core: 3, 2, 1 nm) | ***Peak Separation*** ($\lambda_{VIS}$ - $\lambda_{UV}$) | **FWHM** (UV) | **FWHM** (Vis) |
|---|---|---|---|---|---|
| 4 | 345.5 nm | 615.0 nm | 269.5 nm | 29.9 nm | 49.4 nm |
| 3 | 343.6 nm | 551.3 nm | 207.7 nm | 31.1 nm | 39.6 nm |
| 2 | 341.8 nm | 477.1 nm | 135.3 nm | 30.1 nm | 32.1 nm |

**Table 1**. Spectral parameters of the obtained absorption spectrum for different sizes of CdSe/ZnS CSQDs.

Upon photon absorption, a molecule is typically excited to a higher vibrational level of the first excited electronic state. It subsequently undergoes rapid vibrational relaxation and geometric reorganization, reaching the lowest vibrational level. Fluorescence emission generally occurs from this relaxed state to various vibrational levels of the electronic ground state. Due to energy loss during these non-radiative relaxation processes, the emitted photon possesses lower energy (longer wavelength) compared to the absorbed photon known as the Stokes shift. Solvent interactions, as previously discussed, can enhance or modulate the Stokes shift by contributing to energy dissipation through solvent reorganization and by altering the electronic energy landscape of the fluorophore[43]. Nevertheless, the intrinsic vibrational relaxation within the molecule alone is sufficient to produce a Stokes shift, even in the absence of a solvent[44]. Consequently, a single fluorescent molecule in vacuum, when photoexcited, will still exhibit a Stokes shift in its fluorescence emission spectrum[45].

We obtained the emission spectra of the employed single CSQDs by a dipole emitter model based on the resulting ACS spectra of our FDTD simulation. Due to the Stokes shift, the emission peaks redshift from the absorption peaks of CdSe core, therefore, we added the Stokes shift values of $\Delta\lambda$ = 15.4, 17.0 and 17.7 nm for large, mid-size and small CSQDs respectively in compliance with the previously reported data [11,46–50]. The CdSe/ZnS CSQDs are modelled as dipole sources at the obtained the emission wavelengths (**Figure 2a**). Size dependent dipole emission spectra is provided by assigning Lorentzian distributions with peak points at 494.8, 568.3 and 630.4 nm and full width at half maximum (FWHM) linewidths of 27.4, 33.2 and 35.6 nm in compliance with the previously reported emission data[12,51,52]. It is worth noting that the resulting spectral width is narrower than the ACS counterpart due to the excitation from the lowest energy state (**Figure 2a**).

Next, we studied the dipole emission of single CdSe/ZnS CSQD that is placed in proximity to a spherical AgNP and numerically evaluated the critical distances incorporating the complete EM interaction. **Figure 2b** shows the snapshot of the electric field intensity distribution (at $t = 0.39\ ps$) sourced from the dipole emission of large CdSe/ZnS CSQD (large dot, r= 4 nm) and localized around the AgNP. (The video of complete EM interaction can be found at **Supplementary Video 1**) The dipole injects an 4.11 fs long Lorentzian light pulse covering wavelengths from 622.5 to 637.5 with the abovementioned 27.4 nm FWHM linewidth. The dipole radiation sourced from the CSQD is incident on AgNP and create localized E-field intensities of the order of $10^8$ $V^2/m^2$ (**Figure 2b**). Formation of hot spots around the AgNP increases the local density of optical states (LDOS) which is directly related to the rate of spontaneous emission as $\Gamma \propto \rho(\omega, r)$. To be in compliance with Rayleigh scattering limit, we employed a spherical AgNP with a radius of 10 nm around which the localized electric field is strongly enhanced due to LSPPs.

The box monitors positioned around the system quantify the power radiated by the dipole, the power absorbed by proximal structures, and the power emitted into the far field. These measurements are used to derive the radiative and non-radiative (absorptive) decay rates, normalized with respect to the intrinsic decay rate of the dipole in free space. This normalization enables the calculation of quantum yield (**Figure 2c**). While the fluorophores that are close to the AgNP exhibit very low quantum yield due to non-radiative energy transfer, distant particles exhibit values closer to 1 as expected in the free space model that our simulations are based on. Typically, for visible to near-infrared (400 – 800 nm) excitation, the hotspot region extends from the surface of AgNP up to 20 nm as the saturation of quantum yield can be already seen at this value (**Figure 2c**). Unlike purely analytical approaches, our numerical FDTD approach captures near-field effects, hot spots, and coupling phenomena that are often crucial in plasmonics. In the case of anisotropic materials in nonhomogeneous medium—such as CSQDs near AgNP— extending analytical models to these scenarios becomes extremely difficult or impossible due to the mathematical intractability of Maxwell's equations for such systems. **Figure 2d-f** show the distance dependence of radiative and non-radiative decay rates for increasing CSQD radius respectively. Since plasmons decay non-radiatively, the non-radiative-dominant behaviour at short interparticle distances is often related to dissipation in which the energy from plasmon excitation is converted into heat via electron-phonon and phonon-phonon interactions. This process includes Förster Resonant Energy Transfer (FRET) mechanism between the dipole source of fluorescent CdSe/ZnS CSQD as a donor and the spherical AgNP as an acceptor. FRET efficiency is strictly related to the spectral overlap between the donor and the acceptor. In the presented case, the emission peaks and Lorentzian distribution around them spectrally overlaps with AgNP's peak around 400 – 450 nm. For larger CSQDs that emits around 630 nm, the spectral overlap is less which causes a lower FRET efficiency.

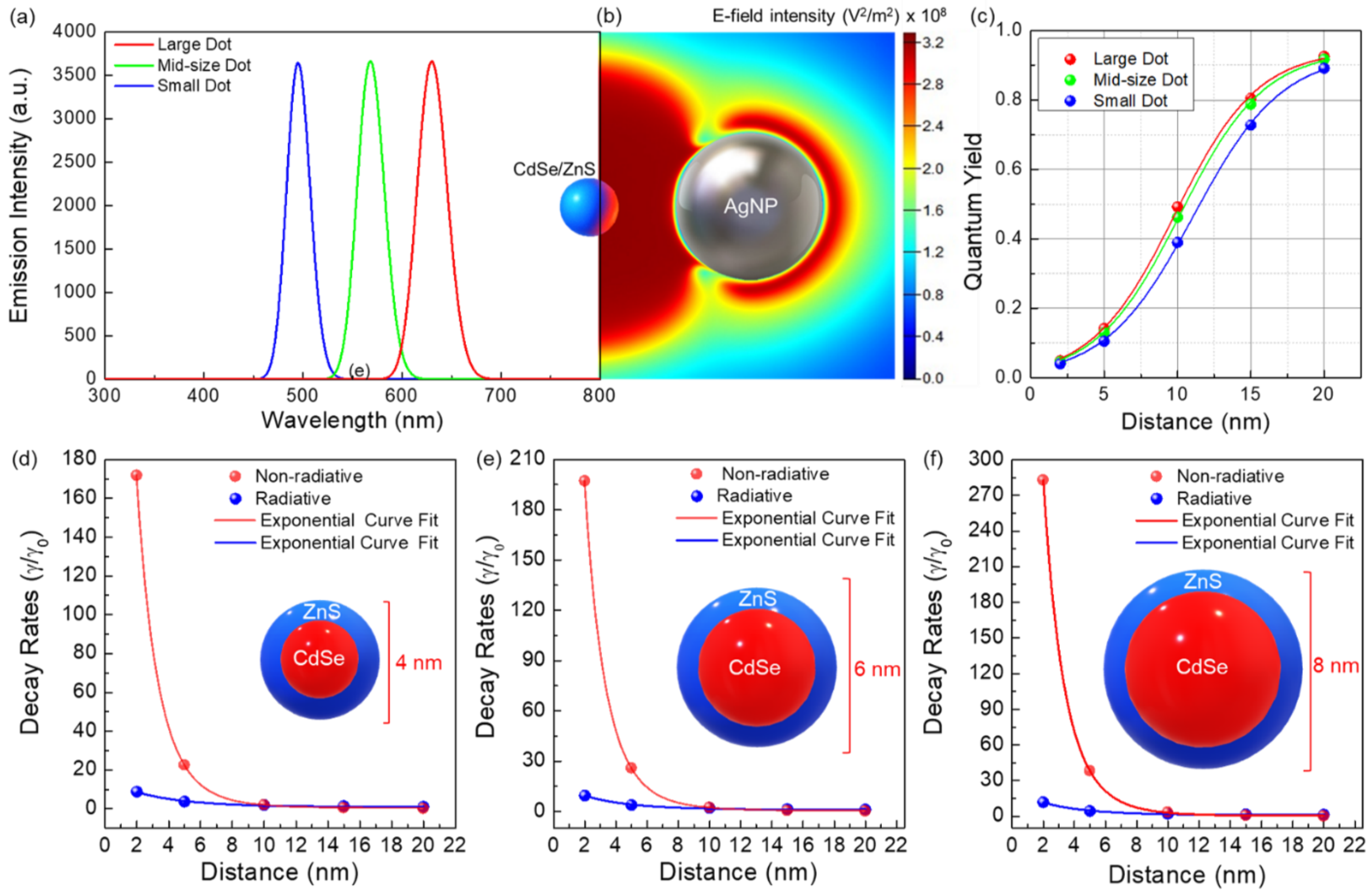

**Figure 2. (a)** PL emission spectra for each size of CdSe/ZnS CSQDs modelled using the corresponding Stokes shifts on the ACS spectra. Lorentzian distribution around peak points at 494.8, 568.3 and 630.4 nm and FWHM linewidths of 27.4, 33.2 and 35.6 nm **(b)** Snapshot of local electric field intensity ($|E_{loc}|^2$) at $t = 0.29$ ps for the presented system of CSQD with 4 nm radius (large dot) and AgNP with 10 nm radius placed at a critical interparticle distance of 10.36 nm. Color bar shows the E-field intensity scale **(c)** Quantum yield with respect to interparticle distance. Values are obtained by the normalization of the decay rates with respect to intrinsic decay rate of the dipole in free space. While closer particles suffer from low quantum yield due to non-radiative decay that suppresses the fluorescence enhancement, distant particles exhibit values closer to 1 as expected in the free space model that our simulations are based. **(d)** Nonradiative and radiative decay rates of CdSe/ZnS CSQDs with respect to interparticle distance for a CSQD of r = 2 nm (small dots), **(e)** for mid-size dots with r = 3 nm and **(f)** for large dots with r = 4 nm. Exponential curve fit in the form of $e^{a+bx+cx^2}$ is applied to determine the critical distances as 12.18 nm 10.73 nm and 10.36 nm for small, mid-size and large dot size respectively.

As shown in the decay rate plots (**Figure 2d-f**), the CSQD's radiative decay rate starts to become equalized to non-radiative decay rate at critical distances. The off-resonance configuration in which there is no complete overlap between AgNP and CdSe/ZnS spectra yields CSQDs with modest fluorescence enhancement after the critical distances. Even under the off-resonance condition, the enhanced local E-field can induce fluorescence enhancement by increasing the excitation rate. Simulating the case in a 3D-FDTD model mainly allows for the implementation of temporal response of the AgNP which includes retardation and multipolar effects that are not taken into account in previous analytical studies reporting distance dependence[36]. Furthermore, the numerical studies that report the same have been mainly focused on the nanoantenna properties of the MNP without explicitly analyzing the interplay between radiative and non-radiative counterpart of the decay for specific fluorophores[25]. In that sense, the critical CSQD-AgNP interparticle distance where the radiative decay is equalized the non-radiative counterpart is obtained by applying a curve fit to the decay rate data for each CSQD size as shown in **Figure 2d-f**. We performed FDTD simulations for each distance step and applied an exponential fit to the extracted data points. The fitting is in the form $e^{a+bx+cx^2}$providing the critical distance values as 10.36, 10.73 and 12.18 nm for large, mid-size and small dots, respectively with a standard error of $\sim \pm 7$ fm for the small and $\sim 0.18$ nm for the midsize and large dot size (Details in Methodology). More interestingly, after the radiative decay is equalized to the non-radiative decay at the critical point, both rates kept decreasing while the non-radiative rate diminishes faster; increasing the difference between the rates that reaches up to 0.96 for the large, 1.06 for the mid-size and 1.07 for the small dot size. These results clearly indicates the benefits of strategic placement of the fluorophore (or plasmonic enhancer) yielding a safe zone for detectable fluorescence enhancement starting from the critical distance value to the limit where the hotspot region ends.

Fluorophores that reside in the hotspot region have much larger probability of undergoing photon absorption and reaching its excited state. Notably, the excitation rate scales with the square of the local electric field intensity ($|E_{loc}|^2$). The detected fluorescence intensity is governed by $I_{fl} \propto |E_{loc}|^2 \cdot \Phi \cdot \eta$ where $\Phi$ is the quantum yield and $\eta$ is the detection efficiency. Keeping the detection efficiency constant we probe the product of local electric field intensity and quantum yield. To quantify the enhancement in the presence of AgNP, we plotted the percentage fluorescence enhancement with respect to distance steps for each CSQD size (**Figure 3).** By comparing the decay rates of the original emission with the boosted radiative decay rate near the AgNP, we created the percentage fluorescence enhancement (FE) spectra within the emission bandwidths of three different dot sizes. The red dashed lines in **Figure 3a-c** show the percentage FE for distance steps shorter than the critical distance where it is suppressed by non-radiative decay, therefore not observable. The FE spectra for distances of 15 nm and 20 nm are shown as dark and light blue solid lines and in **Figure 3a-c** which are longer than the critical distance values thus, observable. The zoomed in spectra showing the observable FE for 15 and 20 nm is shown in **Figure 3d-f**. While the large CSQD shows profound increment in the enhancement reaching up to 160% (**Figure 3c, 3f**), we observed that the effect is less pronounced for the mid-size and the small CSQDs within their emission bandwidth.

To demonstrate the promises of working at critical interparticle distances, we simulate an optically switchable dipole emission of large CdSe/ZnS CSQD (r = 4 nm) at the critical distance (10.36 nm). By irradiating AgNP via collimated light beam shined on the surface (**Figure 4a**), we recorded the effects of the boosted AgNP excitation on the emission spectra of CdSe/ZnS CSQD. **Figure 4a** shows the schematic illustration of the experiment. To specifically investigate the effect of the LSPP enhancement on the fluorescence intensity, a broadband (300 – 800 nm) EM wave pulse is sent to the AgNP surface without any direct contribution to the initial photoluminescence of CdSe/ZnS either from the light source itself or the scattering from the surface. The snapshots of the temporal E-field intensity localization and the resulting LSPP oscillations forming the hotspot regions are shown in **Figure 4b (**Complete interaction can be found at **Supplementary Video 2)**. To obtain maximum fluorescence enhancement, the relative positioning of the incident light and CdSe/ZnS shown in **Figure 4a** are determined by the hotspot regions where the-field intensity localization is maximum (shown as dark red color in normalized color scale).

The recorded radiative and non-radiative decay due to the CdSe/ZnS CSQDs dipole source of under the light excitation of AgNP are shown in **Figure 4c** and **4d**. Performing the optical excitation when the particles are separated by the critical distance value (10.36 nm) allowed us to determine the deviations from initially equalized radiative and non-radiative decay rates. The effect of enhanced field localization in AgNP induces a boost in radiative decay rate (**Figure 4c**) while reducing the rate of non-radiative decay (**Figure 4d**). With the application of a light pulse, we achieved a factor of ~ 2.1 to 2.4 enhancement (solid line in **Figure 4c**) on top of the initial enhancement due to spatial placement at critical distance from AgNP surface (dashed line in **Figure 4c**). Moreover, the same light application also decreases the non-radiative decay rate with a factor of ~2.2 to 2.3, preventing the suppression of radiative decay due to nonradiative decay mechanisms within the PL bandwidth (**Figure 4d**).

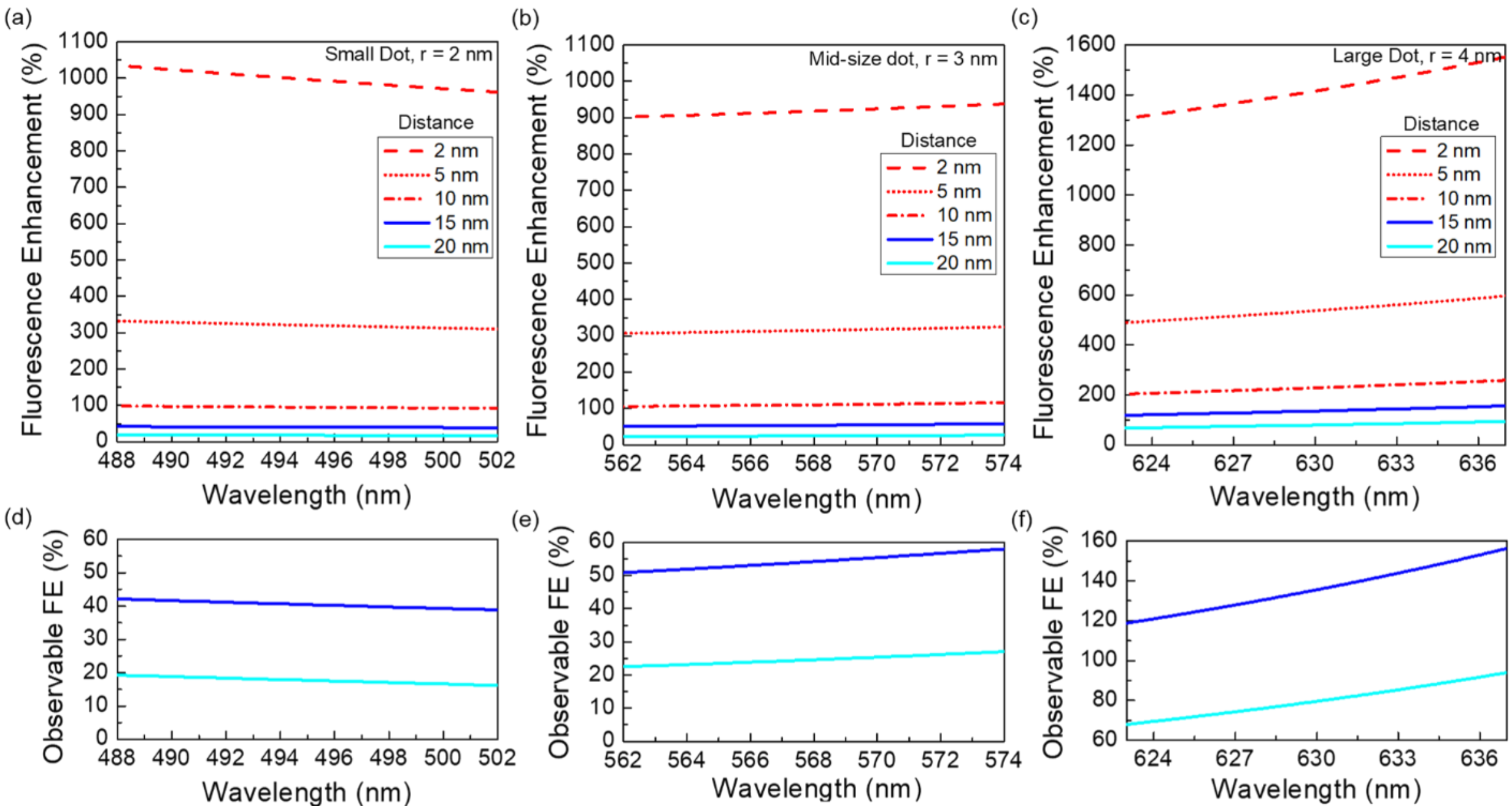

**Figure 3.** Fluorescence enhancement (FE) spectra within the emission bandwidth for **(a)** small **(b)** mid-size **(c)** large CdSe/ZnS CSQD sizes. The red dashed lines showing FE for 2, 5, and 10 nm distance steps that are shorter than the critical distance values. Therefore, FE is not observable for these values (suppressed by non-radiative decay). The observable FE for 15 and 20 nm are shown by solid lines in dark and light blue respectively. The zoomed in spectra showing solely the observable FE for 15 and 20 nm distance values are given in **(d)** for small, **(e)** for mid-size and **(f)** for large CSQD sizes.

The obtained optically switchable fluorescence enhancement at critical distance is shown in **Figure 4e**. According to our model, one can achieve a maximum ~720% fluorescence enhancement due to the increase in the AgNP excitation with the light application, allowing an active control in radiative and non-radiative decay of CdSe/ZnS with light modulation. While the enhancement factor can be controlled by the applied light intensity and wavelength, the total 720% increment can be decomposed into two components as passive enhancement due to spatial proximity to AgNP (dashed line in **Figure 4e**) and the active enhancement due to the light application (solid line in **Figure 4e**). **Table 2** shows the complete set of extracted parameters from our model that considers the fluorophores as physical particles and dipole emitters simultaneously.

The provided FDTD model that reports the results of three-dimensional solutions of Maxwell's curl equations in Rayleigh limit is general enough to be implemented for fluorophores with known refractive index data. The presented free space model can be further adapted for specific device application by modifying the environmental dielectric constant of the target substrate. We emphasize that the reported critical distance and fluorescence enhancement results stand for the Rayleigh limit and increasing the particle size to reach Mie scattering may require an additional analysis. In that sense, we anticipate that our work will serve as a base reference for the photonic systems with small-sized fluorophores and MNPs under broadband illumination.

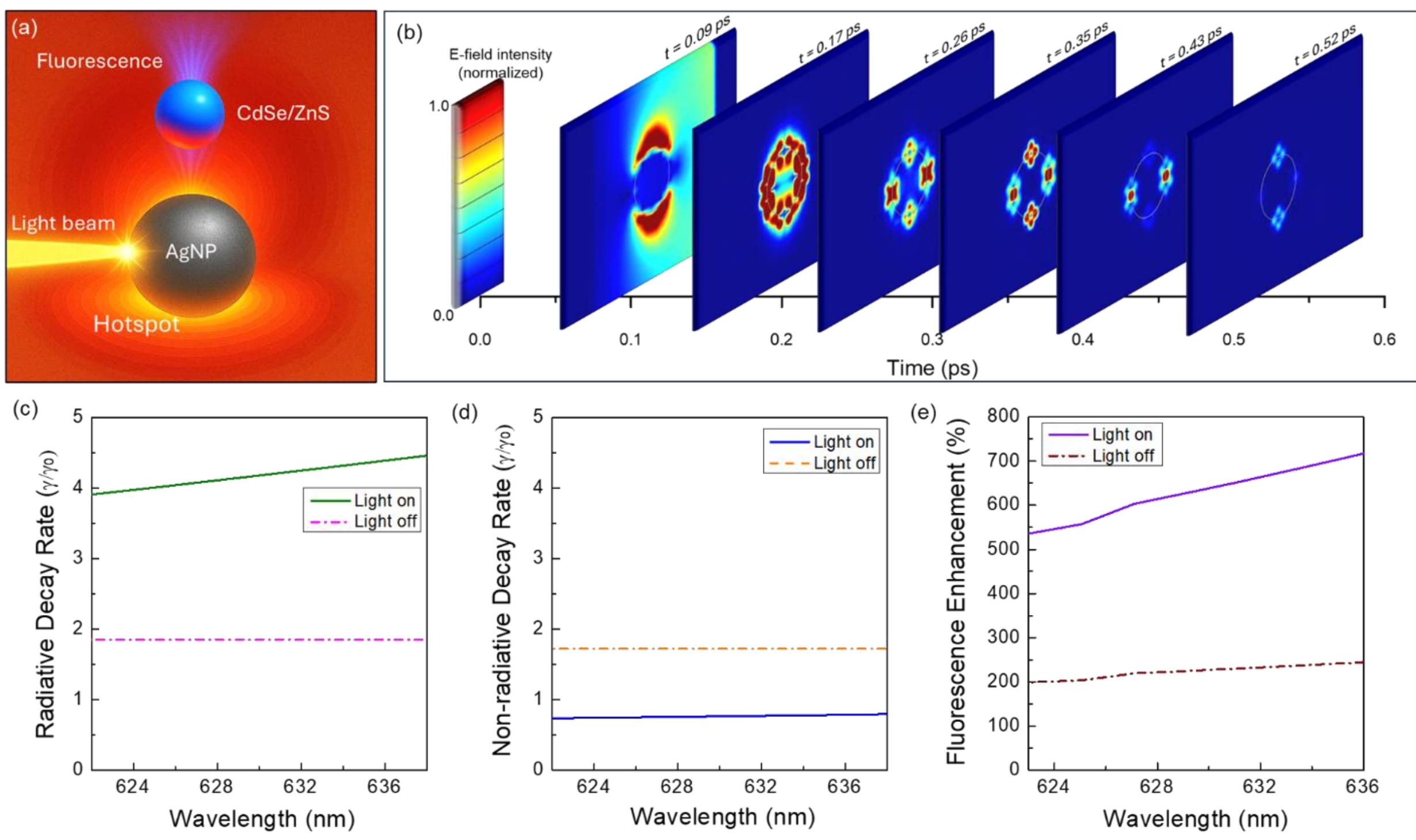

**Figure 4. (a)** Schematic illustration of the optical excitation of AgNP and the resulting fluorescence enhancement from CdSe/ZnS. **(b)** Temporal response showing the E-Field intensity localization on AgNP surface under the injected broadband plane wave pulse (300 – 800 nm). **(c)** Recorded radiative decay rate from the dipole emission of CdSe/ZnS CSQD with and without light pulse injection to AgNP surface. **(d)** Recorded non-radiative decay rate under the same illumination conditions. **(e)** Fluorescence enhancement spectrum showing the optically switchable enhancement behavior reaching up to an observable 720 % enhancement within the emission bandwidth of the investigated CSQD with r = 4 nm (large dot).

| ***CSQD Radius*** **(nm)** | ***Absorption Maxima*** **(nm)** | ***Stokes Shift*** **(nm)** | ***PL Emission Maxima*** **(nm)** | ***Critical Distance*** **(nm)** | **Passive Fluorescence Enhancement** (Spatial positioning) **(%)** | **Active Fluorescence Enhancement** (Optical excitation) **(%)** |
|---|---|---|---|---|---|---|
| 4 | 615.0 | 15.4 | 630.4 | 10.36 | 250 | 720 |
| 3 | 551.3 | 17.0 | 568.3 | 10.73 | 102 | 235 |
| 2 | 477.1 | 17.7 | 494.8 | 12.18 | 66 | 124 |

**Table 2**. Overall parameters and resulting maximum fluorescence enhancement rates due to passive (spatial positioning around critical distance) and active (optical excitation of AgNP) control methods.

## Conclusion and Discussion

After Yee's algorithm to shift electric and magnetic fields in discretized space and time for the solution of Maxwell's equations[53], FDTD method has been applied to many plasmonic and photonic systems with the advancement of computer technology that allows for the discretization of space in sub-wavelength dimensions[54]. Pure analytical methods fall short in calculating the near fields in complex scenarios, therefore, here we implement an FDTD simulation with Yee cell (mesh) size that is much smaller ($\sim$1 nm) than the incident wavelength (300 – 800 nm) that can overcome complex nanostructures with superimposed materials such as CdSe/ZnS CSQDs. By precisely changing the interparticle distance between individual CSQD and AgNP for each CSQD size, we elucidate the EM interactions giving rise to fluorescence enhancement, FRET and non-radiative decay. Our FDTD model provides for the first time critical CSQD-AgNP interparticle distances by simultaneously measuring the radiative and non-radiative decay rates. We extracted the critical distance points where the suppression of radiative decay is no longer effective, yielding observable fluorescence enhancement in the absence of any gain mechanism. This behavior underscores a central principle in plasmon-enhanced fluorescence: maximal brightness is achieved not simply by maximizing the coupling strength, but by balancing radiative enhancement against nonradiative quenching. Our findings are consistent with theoretical models and previous experimental studies which have demonstrated that the highest fluorescence enhancements are observed in systems where the emitter is optimally detuned or positioned to favor radiative decay over nonradiative losses[55].

The implementation of critical distances found in our simulations are demonstrated as an optically switchable fluorescence enhancement in which the AgNP is exposed to broadband illumination. By increasing the local field enhancement on AgNP, we provided an active optical fluorescence intensity control scheme for the large dot (r = 4 nm) yielding up to 720% enhancement without being suppressed by non-radiative decay mechanism. Our model can be generalized to find critical distances and the amount of enhancement in the radiative decay rate for the fluorophores with known refractive index data in the Rayleigh limit where the size properties are much smaller than the wavelength ($d \ll \lambda$). Since the heterogeneity in CSQD-AgNP coupling is often masked in ensemble measurements, we anticipate that our study serves as a benchmark for the development of QD-based light emitting technologies using fluorescence enhancement. Similarly, the provided size dependent ACS and SCS spectra can be implemented in photodetection technologies using the size and positioning of QDs as an active parameter in the absorber layers. Our free space model can be adapted to other media by changing the environmental dielectric constant, which allows for the determination of the critical distances for the device structures or liquid phase dispersions that contain small fluorescent particles.

## Methodology

We simulated CdSe-ZnS CSQDs in free space with and without the presence of closely located AgNP using FDTD technique to solve Maxwell's equations. Maxwell's curl equations are discretized in both time and space:

$$\boldsymbol{\nabla} \times \boldsymbol{E} = -\mu \frac{\partial \boldsymbol{H}}{\partial t}, \qquad \boldsymbol{\nabla} \times \boldsymbol{H} = \epsilon \frac{\partial \boldsymbol{E}}{\partial t} + \boldsymbol{J},$$

Where $\boldsymbol{E}$ and $\boldsymbol{H}$ are electric and magnetic fields, $\mu$ is the permeability, $\epsilon$ is the permittivity and $\boldsymbol{J}$ is the current density.

FDTD enables the simulation of complex interactions between the quantum dot and the plasmonic nanoparticle, capturing both near-field enhancements and far-field scattering/absorption effects. To simulate an open, infinite field and reduce spurious reflections from the field boundaries, a Perfectly

Matched Layer (PML) was implemented in FDTD simulations. The free space model is implemented in 3D FDTD solver (Lumerical, FDTD solutions). The PML was configured to absorb outgoing waves with minimal reflection by gradually increasing the electrical conductivity in the absorber region. This caused the electromagnetic fields to decay exponentially as they approached the boundaries, effectively simulating an infinite field.

To calculate the absorption and scattering cross sections of the CSQDs, the light source (TF/SF) source was emitted from the interior of the PML in the form of a plane wave packet at broadband spectrum (300 – 800 nm). To guarantee numerical stability, the simulation space is discretized in both space and time using the Courant condition. With a constant ZnS shell thickness of 1 nm, three CdSe/ZnS CSQDs with radii 2 nm, 3 nm, and 4 nm are evaluated.

A frequency dependent Lorentzian dielectric function is used for the core and shell materials to characterize the light-matter interaction and the resulting optical response;

$$\epsilon(\omega) = \epsilon_\infty + \frac{f\omega_0^2}{\omega_0^2 - \omega^2 - i\gamma\omega}$$

where, $\epsilon_\infty$ is the background dielectric constant, $f$ is the oscillator strength quantifying the transition strength, $\omega_0$ is the resonance (or natural) frequency corresponding to the electronic transition, $\omega$ is the angular frequency of the incident EM field, $\gamma$ is the damping constant representing losses, and the imaginary unit indicating the phase shift between the driving field and the polarization respons,. The large CSQD has a linewidth of ~ $10^{14}$ Hz, a core Lorentz oscillator strength of 0.15, and a background permittivity of 5.6. The ZnS shell is modeled with a background permittivity of 5.2 and a Lorentzian contribution of 0.5. The ACS and SCS spectra are extracted as:

$$\sigma_{\mathrm{abs}}(\omega) = \frac{\omega}{c\epsilon_0}\mathrm{Im}[\alpha(\omega)], \qquad \sigma_{\mathrm{sca}}(\omega) = \frac{k^4}{6\pi}\mathrm{Re}[\alpha(\omega)],$$

where $k = \frac{\omega}{c}$ is the wavenumber, $\alpha(\omega)$ is the polarizability of the dot, $c$ is the speed of light and $\epsilon_0$ is the permittivity of the free space.

Radiative and non-radiative decay rates as function of CSQD-AgNP interparticle spacing are calculated to assess the decay dynamics. CdSe/ZnS CSQDs are modelled as dipole sources emitting at the corresponding emission frequencies (absorption wavelength + Stokes shift). Their decay rate in free space is expressed as

$$\gamma_0 = \frac{\omega^3|p|^2}{3\pi\varepsilon_0\hbar c^3}$$

where, ω is the angular frequency, $|p|$ is the dipole moment, $\varepsilon_0$ is the vacuum permittivity, $\hbar$ is the Planck's cnstant, and $c$ is the speed of light. To isolate the energy that is emitted as light (radiated power) in the presence of AgNP (experimental data provided by Johnson & Christy [56]), we use far-field power monitors. The radiative decay rate is [57]

$$\gamma_{rad} = \frac{P_{rad}}{P_0}\gamma_0$$

where $P_{rad}$ is the power measured by monitors in the far-field. Total decay rate is then,

$$\gamma_{tot} = \frac{P_{tot}}{P_0}\gamma_0$$

where $P_{tot}$ is the power measured by monitors in the near field. And the non-radiative decay rate is,

$$\gamma_{nonrad} = \gamma_{tot} - \gamma_{rad}$$

Near-field power monitors positioned to cover emitting dipole and AgNP respectively to capture all energy flows, including those modified by the antenna's near-field coupling, which is

$$P_{tot} = \oint_s S.\, dA$$

where S is the Poynting vector. Analogously,

$$P_{rad} = \oint_{S_{far}} S.\, dA$$

is the power calculated by far-field monitors of a chosen surface (simulation domain) such that non-radiative losses are not captured, isolating the radiative decay enhanced by LSPR. In the case of optical excitation of AgNP by light application, the exclusion of the direct intensity contribution to the measurement of dipole emission is guaranteed via transmission monitors placed around the AgNP that shows zero intensity of the light scattered from the AgNP surface and reach to the monitor. Therefore, the recorded fluorescence enhancement of CdSe/ZnS CSQD placed at a critical distance is solely due to the optical excitation of AgNP.

The distance dependence of these rates is quantified by application of exponential curve fit. Standard error in the critical distances are calculated from the applied exponent in the form:

$$y_i(x) = exp(a_i + b_i x + c_i x^2), \qquad i = 1,2$$

where the coefficients $a_i$, $b_i$, and $c_i$ are obtained from independent fits (see **Table 3**).

| ***CSQD Radius* (nm)** | **Radiative "a"** | **Radiative "b"** | **Radiative "c"** | **Non-radiative "a"** | **Non-radiative "b"** | **Non-radiative "c"** |
|---|---|---|---|---|---|---|
| 4 | 2.77905 | -0.32861 | 0.0096 | 6.71346 | -0.82581 | 0.02106 |
| 3 | 2.88186 | -0.34493 | 0.01075 | 6.85067 | -0.82569 | 0.02112 |
| 2 | 3.17591 | -0.38235 | 0.01211 | 7.18201 | -0.80915 | 0.02014 |

**Table 3**. Fitting parameters for the exponential function $y_i(x) = exp(a_i + b_i x + c_i x^2)$ for each CSQD size.

The critical (or intersection) point $x^*$ is defined by the condition

$$y_1(x^*) = y_2(x^*).$$

Defining a difference function at the intersection,

$$F(x) = y_1(x) - y_2(x), \text{ with } F(x^*) = 0.$$

We assume that near the critical point the measured $y$ -values are subject to a typical uncertainty $\delta y$. If the uncertainties for $y_1$ and $y_2$ are approximately equal and independent, the uncertainty in the difference is given by

$$\delta F \approx \sqrt{(\delta y_1)^2 + (\delta y_2)^2} \approx \sqrt{2}\delta y.$$

Expanding $F(x)$ in a first-order Taylor series about $x^*$:

$$F(x^* + \delta x) \approx F(x^*) + F'(x^*)\delta x.$$

At the intersection point,

$$\delta F \approx F'(x^*)\delta x.$$

Rearranging,

$$\delta x \approx \frac{\delta F}{| F'(x*) |} \approx \frac{\sqrt{2}\delta y}{| F'(x*) |}$$

where, $F'(x^*)$ is the derivative of the difference function at $x = x^*$. To compute $F'(x^*)$, note that each curve is given by $y_i(x) = exp(a_i + b_i x + c_i x^2)$ and its derivative with respect to x is $y_i'(x) = y_i(x)(b_i + c_i x)$. Thus, the derivative of the difference function is

$$F(x^*) = y_1'(x) - y_2'(x) = y_1(x)(b_1 + c_1 x) - y_2(x)(b_2 + c_2 x)$$

Evaluating at the critical point, $x = x^*$. the uncertainty in critical distances are found using:

$$\delta x \approx \frac{\sqrt{2}\delta y}{| y_1(x)(b_1 + c_1 x) - y_2(x)(b_2 + c_2 x) |}$$

**Acknowledgments**

E.O.P acknowledges the support from TUBITAK (222N308, CHIST-ERA) and Bilkent University-TUBITAK BILGEM Consultancy Call for Research Proposals for the 2024-2025 academic year.

**Author Contributions**

A.G ran the simulations, performed the calculations and extracted the the raw data forming the basis of the shown results and contributed to the formation of the manuscript. E.O.P conceptualized and supervised the idea and majorly contributed to the formation of the manuscript.

**Competing interests**

The authors declare no competing interests.

**Data availability**

All the data needed to evaluate the shown results is present within the paper. Additional data related to the paper that support the findings of this study are available from the corresponding author upon reasonable requests.

**References**

[1] J. Henzie, J. Lee, M. H. Lee, W. Hasan, T. W. Odom, *Annu. Rev. Phys. Chem.* **2009**, *60*, *147-165*.

[2] A. V. Zayats, I. I. Smolyaninov, in *J. Opt. A: Pure Appl. Op.*, **2003,** *5, S16*.

[3] J. K. So, G. H. Yuan, C. Soci, N. I. Zheludev, *Appl. Phys. Lett.* **2020**, *117*.

[4] F. Tam, G. P. Goodrich, B. R. Johnson, N. J. Halas, *Nano Lett.* **2007**, *7*, *496-501*.

[5] T. N. A. Mai, M. S. Hossain, N. M. Nguyen, Y. Chen, C. Chen, X. Xu, Q. T. Trinh, T. Dinh, T. T. Tran, *Adv. Funct. Mater.* **2025**, 2500714.

[6] S. Singh, S. A. Catledge, *J. Appl. Phys.* **2013**, *113(4)*.

[7] T. V. Shahbazyan, *Phys. Rev. Lett.* **2016**, *117*, *207401*.

[8] X. Peng, L. Manna, W. Yang, J. Wickham, E. Scher, A. Kadavanich, A. P. Alivisatos, *Nature 2000 404:6773* **2000**, *404*, 59.

[9] Y. Fu, D. Kim, W. Jiang, W. Yin, T. K. Ahn, H. Chae, *RSC Adv.* **2017**, *7*, 40866.

[10] V. L. Colvin, M. C. Schlamp, A. P. Alivisatos, *Nature 1994 370:6488* **1994**, *370*, 354.

[11] B. O. Dabbousi, J. Rodriguez-Viejo, F. V. Mikulec, J. R. Heine, H. Mattoussi, R. Ober, K. F. Jensen, M. G. Bawendi, *J. Phys. Chem. B* **1997**, *101*, 9463.

[12] J. Jasieniak, L. Smith, J. Van Embden, P. Mulvaney, M. Califano, *J. Phys. Chem. C* **2009**, *113*, 19468.

[13] A. M. Smith, S. Nie, *Acc. Chem. Res.* **2010**, *43*, *190-200*.

[14] S. Kasap and P. Capper, *Springer Handbook of Electronic and Photonic Materials*, Springer New York, US, **2017**.

[15] I. L. Medintz, H. T. Uyeda, E. R. Goldman, H. Mattoussi, *Nat. Mater. 2005 4:6* **2005**, *4*, 435.

[16] Y. M. Huang, K. J. Singh, A. C. Liu, C. C. Lin, Z. Chen, K. Wang, Y. Lin, Z. Liu, T. Wu, H. C. Kuo, *Nanomaterials* **2020**, *10*, 1327.

[17] K. H. Ibnaouf, S. Prasad, A. Hamdan, M. Alsalhi, A. S. Aldwayyan, M. B. Zaman, V. Masilamani, *Spectrochim. Acta. A Mo.l Biomol. Spectrosc.* **2014**, *121*, *339-345*.

[18] A. Polovitsyn, Z. Dang, J. L. Movilla, B. Martín-García, A. H. Khan, G. H. V. Bertrand, R. Brescia, I. Moreels, *Chem. Mater.* **2017**, *29*, 5671.

[19] J. Franck, *Trans. Faraday Soc.***1926**, *21*, *536-542*.

[20] E. Condon, *Phys. Rev.* **1926**, *28*, *1182*.

[21] E. L. Mertz, V. A. Tikhomirov, L. I. Krishtalik, *J Phys. Chem. A* **1997**, *101, 3433-3442*.

[22] A. F. Koenderink, A. Alù, A. Polman, *Science (1979)* **2015**, *348*, 516.

[23] E. M. Purcell, *Phys. Rev.* **1946**, *69*.

[24] A. E. Krasnok, A. P. Slobozhanyuk, C. R. Simovski, S. A. Tretyakov, A. N. Poddubny, A. E. Miroshnichenko, Y. S. Kivshar, P. A. Belov, *Sci. Rep.* **2015**, *5*, *12956*.

[25] J. L. Montaño-Priede, M. Zapata-Herrera, R. Esteban, N. Zabala, J. Aizpurua, *Nanophotonics* **2024**, *13*, 4771.

[26] J. B. Khurgin, G. Sun, *Appl. Phys. Lett.* **2011**, *99*, 211106.

[27] Y. Yang, A. Dev, I. Sychugov, C. Hägglund, S. L. Zhang, *J Phys. Chem.* **2023**, *14*, 2339.

[28] K. Zhang, D. Zhou, J. Qiu, Z. Long, R. Zhu, Q. Wang, J. Lai, H. Wu, C. Zhu, *J. Am. Ceram. Soc.* **2020**, *103*, 2463.

[29] S. Xiao, V. P. Drachev, A. V. Kildishev, X. Ni, U. K. Chettiar, H. K. Yuan, V. M. Shalaev, *Nature 2010 466:7307* **2010**, *466*, 735.

[30] D. Güney, T. Koschny, C. M. Soukoulis, *Phys. Rev. B* **2009**, *80*, 125129.

[31] J. B. Khurgin, A. Boltasseva, *MRS Bull* **2012**, *37*, 768.

[32] M. Nirmal, B. O. Dabbousi, M. G. Bawendi, J. J. Macklin, J. K. Trautman, T. D. Harris, L. E. Brus, *Nature* **1996**, *383*, 802.

[33] T. Takada, C. Y. Li, J. Y. Tseng, J. D. Mackenzie, *J Solgel. Sci. Technol.* **1994**, *1*, 123.

[34] H. Jin, B. Baek, D. Kim, F. Wu, J. D. Batteas, J. Cheon, D. H. Son, *Nano Lett.* **2017**, *17*, 7471.

[35] Z. Chen, A. Manian, A. Widmer-Cooper, S. P. Russo, P. Mulvaney, *Chem. Rev.* **2025**, *125*, 4359.

[36] R. Carminati, J. J. Greffet, C. Henkel, J. M. Vigoureux, *Opt. Commun.* **2006**, *261*, *368-375*.

[37] M. Thomas, J. J. Greffet, R. Carminati, J. R. Arias-Gonzalez, *Appl. Phys. Lett.* **2004**, *85*, 3863.

[38] L. A. Blanco, F. J. García De Abajo, *Phys. Rev. B* **2004**, *69*, 205414.

[39] L. Novotny, *Appl. Phys. Lett.* **1996**, *69*, 3806.

[40] V. V. Klimov, M. Ducloy, V. S. Letokhov, *Quantum Electron.* **2001**, *31*, 569.

[41] K. Umashankar, A. Taflove, *IEEE Trans. Electromagn. Compat.* **1982**, *EMC-24*, 397.

[42] D. E. Merewether, R. Fisher, F. W. Smith, *IEEE Trans. Nucl. Sci.* **1980**, *27*, 1829.

[43] K. G. Fleming, *Encyclopedia of Spectroscopy and Spectrometry, Second Edition,* Academic Press, Oxford, UK, **2010**.

[44] M. H. W. Stopel, C. Blum, V. Subramaniam, *J Phys. Chem.* **2014**, *5*, 3259.

[45] A. Bagga, P. K. Chattopadhyay, S. Ghosh, *Phys. Rev. B* **2006**, *74*, 035341.

[46] M. R. Karim, M. Balaban, H. Unlü, *Adv. Mater. Sci. Eng.* **2019,** *2019, 3764395*.

[47] A. L. Efros, M. Rosen, M. Kuno, M. Nirmal, D. Norris, M. Bawendi, *Phys. Rev. B* **1996**, *54*, *4843*.

[48] A. M. Smith, S. Nie, *Analyst* **2004**, *129, 672-677*.

[49] T. J. Liptay, L. F. Marshall, P. S. Rao, R. J. Ram, M. G. Bawendi, *Phys. Rev. B* **2007**, *76*, *155314*.

[50] S. Mathew, A. D. Saran, B. Singh Bhardwaj, S. Ani Joseph, P. Radhakrishnan, V. P. N. Nampoori, C. P. G. Vallabhan, J. R. Bellare, in *J Appl. Phys.* **2012**, *111, 74312.*

[51] D. V. Talapin, A. L. Rogach, A. Kornowski, M. Haase, H. Weller, *Nano Lett.* **2001**, *1*, 207.

[52] X. Peng, M. C. Schlamp, A. V. Kadavanich, A. P. Alivisatos, *J Am. Chem. Soc.* **1997**, *119*, 7019.

[53] K. S. Yee, *IEEE Trans. Antennas Propag.* **1966**, *14*, 302.

[54] A. C. Lesina, P. Berini, A. Vaccari, L. Ramunno, *Opt. Express* **2015**, *23*, 10481.

[55] P. Anger, P. Bharadwaj, L. Novotny, *Phys. Rev. Lett.* **2006**, *96*, 113002.

[56] P. B. Johnson, R. W. Christy, *Phys. Rev. B* **1972**, *6*, 4370.

[57] L. Novotny, B. Hecht, *Principles of Nano-Optics,* Cambridge University Press*,* Cambridge, UK, **2012**.